\documentclass[aps,prb,reprint,superscriptaddress]{revtex4-1}

\usepackage{graphicx}

\begin{document}


\title{A sensitive calorimetric technique to study energy (heat) exchange at the nano-scale}



\author{Luca Basta}
\email[]{luca.basta1@sns.it}
\affiliation{NEST, Istituto Nanoscienze-CNR and Scuola Normale Superiore, Piazza S. Silvestro 12, 56127 Pisa, Italy}

\author{Stefano Veronesi}
\email[]{stefano.veronesi@nano.cnr.it}
\affiliation{NEST, Istituto Nanoscienze-CNR and Scuola Normale Superiore, Piazza S. Silvestro 12, 56127 Pisa, Italy}

\author{Yuya Murata}
\affiliation{NEST, Istituto Nanoscienze-CNR and Scuola Normale Superiore, Piazza S. Silvestro 12, 56127 Pisa, Italy}

\author{Zo\'{e} Dubois}
\affiliation{NEST, Istituto Nanoscienze-CNR and Scuola Normale Superiore, Piazza S. Silvestro 12, 56127 Pisa, Italy}

\author{Neeraj Mishra}
\affiliation{Center for Nanotechnology Innovation @ NEST, Istituto Italiano di Tecnologia, Piazza S. Silvestro 12, 56127 Pisa, Italy}
\affiliation{Graphene Labs, Istituto Italiano di Tecnologia, Via Morego 30, 16163 Genova, Italy}

\author{Filippo Fabbri}
\affiliation{Center for Nanotechnology Innovation @ NEST, Istituto Italiano di Tecnologia, Piazza S. Silvestro 12, 56127 Pisa, Italy}
\affiliation{Graphene Labs, Istituto Italiano di Tecnologia, Via Morego 30, 16163 Genova, Italy}

\author{Camilla Coletti}
\affiliation{Center for Nanotechnology Innovation @ NEST, Istituto Italiano di Tecnologia, Piazza S. Silvestro 12, 56127 Pisa, Italy}
\affiliation{Graphene Labs, Istituto Italiano di Tecnologia, Via Morego 30, 16163 Genova, Italy}

\author{Stefan Heun}
\affiliation{NEST, Istituto Nanoscienze-CNR and Scuola Normale Superiore, Piazza S. Silvestro 12, 56127 Pisa, Italy}


\date{\today}

\begin{abstract}
Every time a chemical reaction occurs, an energy exchange between reactants and environment exists, which is defined as the enthalpy of the reaction. In the last decades, research has resulted in an increasing number of devices at the micro- or nano-scale. Sensors, catalyzers, and energy storage systems are more and more developed as nano-devices which represent the building blocks for commercial "macroscopic" objects. A general method for the direct evaluation of the energy balance of such systems is not available at present. Calorimetry is a powerful tool to investigate energy exchange, but it usually needs macroscopic sample quantities. Here we report on the development of an original experimental setup able to detect temperature variations as low as 10~mK in a sample of $\sim 10$~ng using a thermometer device having physical dimensions of $5 \times 5$~mm$^2$. The technique has been utilized to measure the enthalpy release during the adsorption process of H$_2$ on a titanium decorated monolayer graphene. The sensitivity of these thermometers is high enough to detect a hydrogen uptake of $\sim$ 10$^{-10}$ moles, corresponding to $\sim$ 0.2 ng, with an enthalpy release of about 23 $\mu$J. The experimental setup allows, in perspective, the scalability to even smaller sizes. \\
\end{abstract}


\maketitle


\section{Introduction}

The measurement of energy/heat exchange between a system and its environment is a very important issue in many chemical and physical problems. An energy flux accompanies any evolution of a system, giving invaluable information on the physicochemical processes underlying the evolution. Calorimetry is defined as the \textsl{measurement of heat}. It has been widely utilized to investigate the properties of matter,\cite{Brown03,Schick16} particularly in the presence of processes affecting the sample structure or inducing changes in the sample's thermodynamic conditions (chemical reactions, phase transitions, etc.). In fact, whenever a temperature gradient exists within a system, or whenever two systems at different temperatures are brought into contact, energy is transferred \textit{via} \textsl{heat transfer}. The analysis of the energy exchange, performed with several calorimetric techniques, has played an important role in the study of the liquid and the solid state.\cite{Haines16}

In general, every time a chemical reaction occurs, its energy balance can be positive or negative. In the former case, energy from the environment is needed, and the reaction is called \textsl{endothermic}. On the contrary, when an \textsl{exothermic} reaction occurs, heat is released to the environment.\cite{Atkins10} Hence, the analysis of energy exchange gives useful information about the physical-chemical properties of a sample. While the absolute amount of energy in a chemical system is difficult to measure or to calculate, the \textsl{enthalpy variation} $\Delta H$ is much easier to work with. The enthalpy variation consists in the change in internal energy $\Delta U$ of the system plus the work $L$ needed to change the system's volume $V$:\cite{Brown03}
\begin{equation}\label{DeltaH1}
\Delta H=\Delta U+L=C_p\cdot\Delta T+V\cdot\Delta P=\delta Q+V\cdot\Delta P
\end{equation} 
where $C_p$ is the heat capacity at constant pressure, $\Delta T$ and $\Delta P$ are temperature and pressure variation, respectively, and $\delta Q=C_p\cdot\Delta T$. For processes under constant pressure $(\Delta P=0)$, $\Delta H$ is the thermodynamic quantity equivalent to the total heat exchanged by the system in endothermic or exothermic reactions: $\Delta H=\delta Q$. Therefore, the enthalpy variation is a useful quantity for studying energy exchange in calorimetric measurements.

So far, calorimetry has, as main limitation, the requirement of macroscopic samples having mass of the order of the \textsl{milligram}, more commonly in the range of $10-100$ mg,\cite{PerkinElmer17} and a limited sensitivity in energy ($E_{min}$ > 1 mJ). Small bolometers allow a better energy resolution, but are limited to the pulsed regime. In this case, the detected average energy can be smaller, but energy is concentrated in short bursts having duration from $\mu$s down to the fs  range. Recently, thermometric techniques based on quantum dots have been adapted, in special cases, for use on the microscopic scale. For instance, they have been utilized for the analysis of thermoelectric devices.\cite{Hoffmann08} Such thermometers, however, can operate only at low temperature (below a few Kelvin). They measure the electron temperature which could be different from the lattice temperature, especially at ultra--low temperatures, and have a sensitivity of the order of 1 mK \cite{Mavalankar13} (see the reference for a detailed discussion on low temperature thermometry).

To give an example of the limited sensitivity of calorimetry, let us consider the adsorption of hydrogen on monolayer graphene. Assuming the DOE prescription \cite{DOEtargets} on gravimetric density (5.5 wt.\% by 2020) and the specific surface area of graphene ($\sim 2600$~m$^2$/g),\cite{Bonaccorso15} the adsorption of 10~mg of H$_2$ on MLG would need $\sim 450$~m$^2$ of graphene (the area of a basketball court). Thermal Desorption Spectroscopy (TDS), on the other hand, is able to detect a very small amount of gas, with an overall sensitivity around $\sim 10^{-11}$ moles.\cite{DeJong90,Denisov01} However, TDS is a destructive technique, in the sense that the desorption of the adsorbed gas is required in order to evaluate its amount. Moreover, in the presence of a desorption barrier, the binding energy obtained through TDS would include it, while the calorimetric technique would not be affected. In summary, a general method to directly evaluate the energy balance of nano-systems in a wide temperature range is not available at present.

Here, we present an original development of the calorimetric technique, which extends the application of calorimetry to microscopic devices. In particular, we demonstrate the detection of the enthalpy released during the adsorption process of $\sim 10^{-10}$~moles of H$_2$ on Ti--functionalized monolayer graphene (Ti--MLG). Our experimental setup is able to detect a hydrogen amount as low as $\sim 0.2$~ng, using a thermometer device having physical dimensions of $\sim 5 \times 5$~mm$^2$. The thermometer we present here has been utilized between 300 and 550 K, but can operate in a larger temperature scale (from 50 to 900 K, where the gold temperature coefficient is constant), achieving a Noise Equivalent Temperature Difference (NETD) of 10 mK. Our work provides a new experimental tool in order to directly measure the heat released during the loading of the sample (hydrogen, in our case, but other gases or materials would work as well), from which we can deduce the amount of  adsorbed material. The main advantages of this calorimetric technique rely on the scalability of sample dimensions and in the measurement of the adsorption energy. Scalability can be exploited towards lower sample dimension (using SiN membranes, for example) while the energy evaluation is non destructive because the technique does not require desorption. 

We choose the Ti--MLG system in order to demonstrate the validity of our calorimetric technique, because of the extensive investigations on solid-state graphene-based devices for their application in the hydrogen storage field.\cite{Kubas09,Tozzini11,Patchkovskii05,Miura03,Goler13,Casolo08,Stojkovic03,Elias09,Ferro08,Burress10,Panella05} In fact, while hydrogen is currently considered one of the most promising clean fuels,\cite{Shell17} the problem of finding an efficient and safe hydrogen storage system is still unsolved.\cite{Zuttel02,Irani02,Sherif97,Principi09,Tozzini13} In particular, metal-functionalized MLG has shown interesting properties such as stable hydrogen storage,\cite{Ataca08,Liu10,Durgun08,Mashoff13} fast adsorption and desorption,\cite{Mashoff13} desorption energies in a useful range for practical applications,\cite{Liu10,Ataca08,Beheshti11,Takahashi16} and in perspective high storage density.\cite{Liu10,Ataca08,Parambhath11,Mashoff15,Beheshti11}

In our study, we realized a sensitive gold film thermometer, which acts as temperature probe and sample holder for the Ti--MLG. After a careful characterization and calibration of the thermometer, a temperature increase (in the range 0.05--0.25 K) occurring in several hydrogen loading experiments has been measured. These results represent the first direct measurements of enthalpy ($\mathrm{H_r}$) released during the hydrogen loading process in functionalized graphene (which is an exothermic process). The corresponding values have been cross-checked through TDS, which allows to extract the loaded hydrogen amount and the related binding energy, showing a very good agreement.

\section{Results and discussion}

\subsection{Experimental setup}

All experiments are performed in an Ultra-High Vacuum (UHV) environment, inside a STM (RHK Technology) chamber. The operations are controlled \textit{via} PC, using specific LabVIEW software which allows to perform heating ramps, data acquisition, and TDS. Our thermal probe is a gold film thermometer (see Figure S1 for details) which allows to evaluate the enthalpy release during the $\mathrm{H_2}$ adsorption. The electrical resistance of the Au film increases with temperature, following the linear relation $R(T)=R_0\left[1+\alpha\left(T-T_0\right)\right]$ where $R_0$ is the resistance at the reference temperature $T_0$ (room temperature in our case) and $\alpha$ is the resistance temperature coefficient.\cite{Serway12} Therefore, a temperature increase of the sensor leads to a resistance increase of the Au thermometer that can be measured. The film resistance is measured with a volt-amperometric technique, using a Wheatstone Bridge cascaded to a  high quality PreAmplifier (for details, see Figure S2). The setup has been improved up to a sensitivity of $\mathrm{\Delta R} \sim 0.03$~$\mathrm{m\Omega}$, corresponding to $\mathrm{\Delta T} \sim 0.01$~K.

\subsection{Thermometer calibration}

In order to properly calibrate the thermometer we performed several heating cycles, recording Resistance vs Temperature curves. In each cycle the sensor is heated up from room temperature (RT) to $\sim 600$~K, using a thermocouple as reference (temperature) sensor. From these curves we obtain the initial resistance temperature coefficient $\alpha_i$ of the sensor (\textit{via} a linear fit) and the electrical resistivity $\rho_i$ ($\rho=R\cdot A/l$, where $R$ is the sample's resistance, $A$ its cross--sectional area, and $l$ its diagonal length). In order to characterize the repeatability and accuracy of the thermometers we have performed the calibration on several samples. Moreover, the same sample has been used in different calibration runs, after being mounted on the sample holder three times, and after staying for several days in the UHV chamber. The resulting data (see Figure S3) show a good overall agreement within $\sim 15$\%. 

\begin{figure}[t]
   \includegraphics[width=\columnwidth]{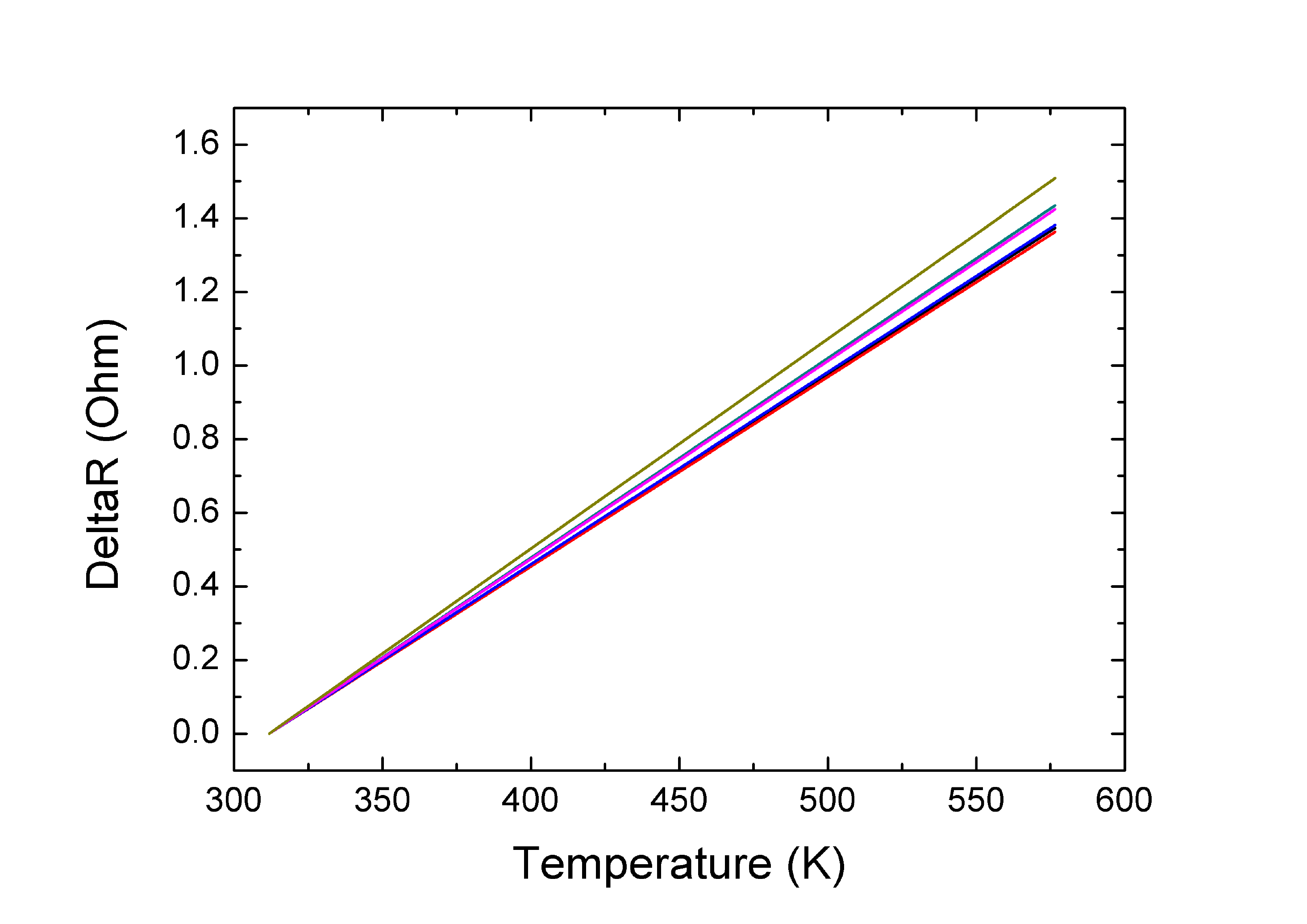}
   \caption{\label{DeltaRG3}Resistance variation $\left(\Delta R=R\left(T\right)-R_0\right)$ vs. temperature for six different heating experiments on the same Au+Ti--MLG sensor.}
\end{figure}

\begin{table}[ht]
 \begin{center}
 	\small
 	\caption{Temperature coefficient of resistance ($\alpha_f$) for each of the heating ramps presented in Figure~\ref{DeltaRG3}. The average value is $\alpha_f = (1.62 \pm 0.05) \cdot 10^{-3}$~$\mathrm{K^{-1}}$.}
 	\label{alphaG3}
     \begin{tabular*}{0.48\textwidth}{@{\extracolsep{\fill}}cc}
       \hline
       \textbf{Ramp} & \textbf{$\alpha_f$ ($\mathrm{K^{-1}}$)} \\
       \hline
       1 & $(1.57\pm0.01) \cdot10^{-3}$ \\ 
       2 & $(1.56\pm0.01) \cdot10^{-3}$ \\
   	   3 & $(1.58\pm0.02) \cdot10^{-3}$ \\
	   4 & $(1.64\pm0.02) \cdot10^{-3}$ \\
	   5 & $(1.63\pm0.02) \cdot10^{-3}$ \\
   	   6 & $(1.72\pm0.02) \cdot10^{-3}$ \\
   	   \hline
     \end{tabular*}
 \end{center}
\end{table}

After this initial thermometer characterization, we transfer a MLG (grown \textit{via} CVD on a copper substrate) on the gold surface of the sensor. The graphene quality and thickness have been checked \textit{via} Raman spectroscopy, which confirmed that the sample is MLG, with no considerable presence of defects (for more details see Figure S4). A new calibration run was performed on each Au+MLG sample. Finally, titanium was deposited on the graphene to functionalize it, \textit{via} evaporation \textit{in situ} inside the UHV chamber (base pressure $\sim 10^{-10}$~mbar). It is well known that Ti on graphene tends to cluster, and for larger amounts of Ti it forms bigger islands. This is confirmed by previous studies in literature \cite{Mashoff13,Mashoff15} and our calibration of the evaporator \textit{via} scanning tunneling microscopy (STM) (see Figures S5 and S6 and Table S2 for details). Therefore, in order to ensure a 100\% coverage of the graphene surface with Ti, we deposited a minimum of 6.5 ML of Ti (1 ML = 1.32$\cdot10^{15}$ Ti atoms per cm$^{2}$).\cite{Mashoff15} After deposition of Ti we repeated the thermometer calibration, obtaining the final calibration parameters $\alpha_f$ and $\rho_f$. Figure~\ref{DeltaRG3} shows some $R$ vs $T$ heating ramps measured with the same sensor. As expected, the curves are linear and reproducible. Table~\ref{alphaG3} lists the $\alpha_f$ values for each ramp. The values of $\alpha_f$ vary within a range of 5\%, showing a good repeatability of the measurements. The same procedure has been repeated on different samples, and the values are in good agreement (see Table S2 for the complete list of parameters for 5 different samples).

\subsection{STM analysis}

\begin{figure*}[t]
   \begin{center}
      \includegraphics[width=0.32\textwidth]{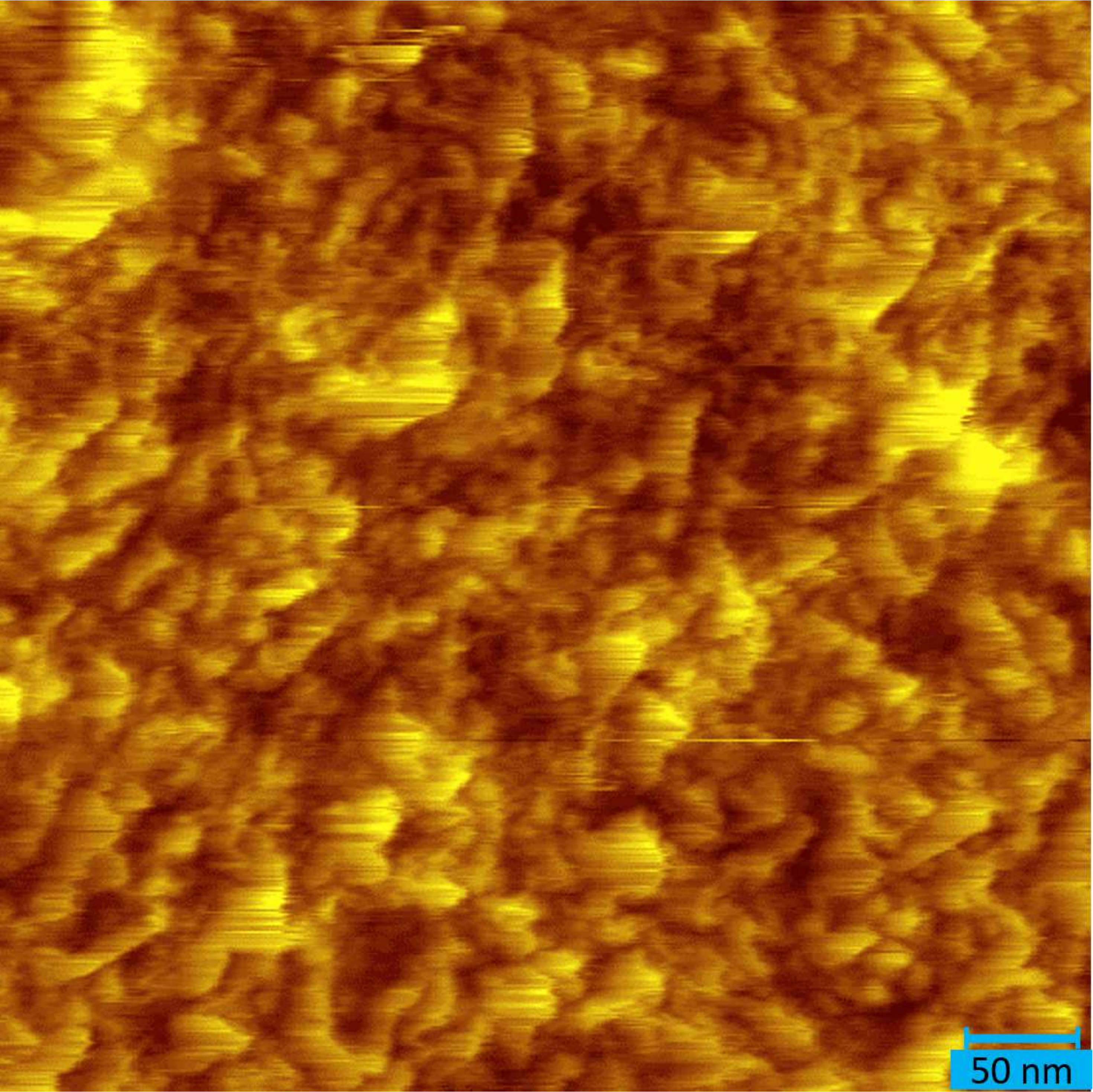}
      \includegraphics[width=0.32\textwidth]{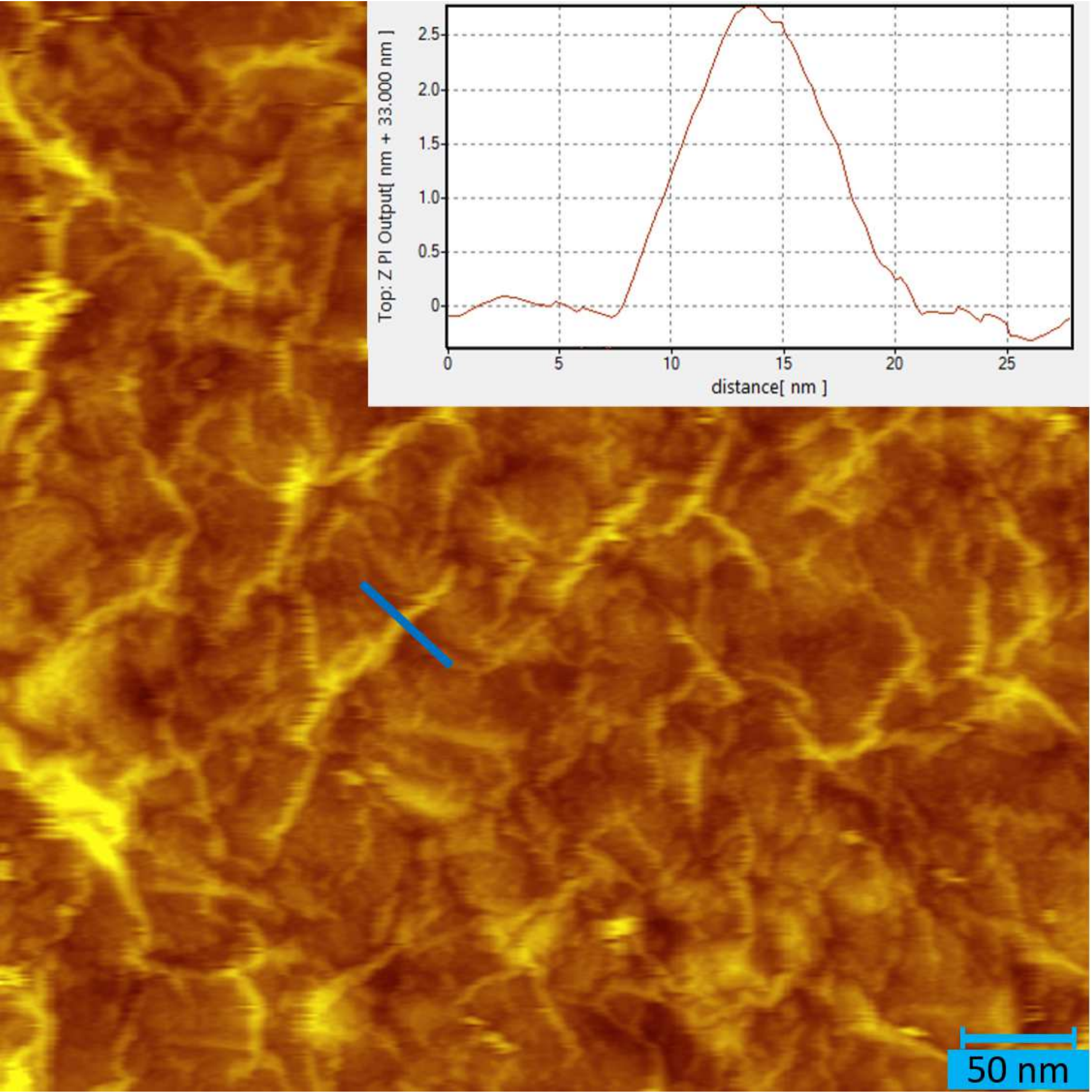}
      \includegraphics[width=0.32\textwidth]{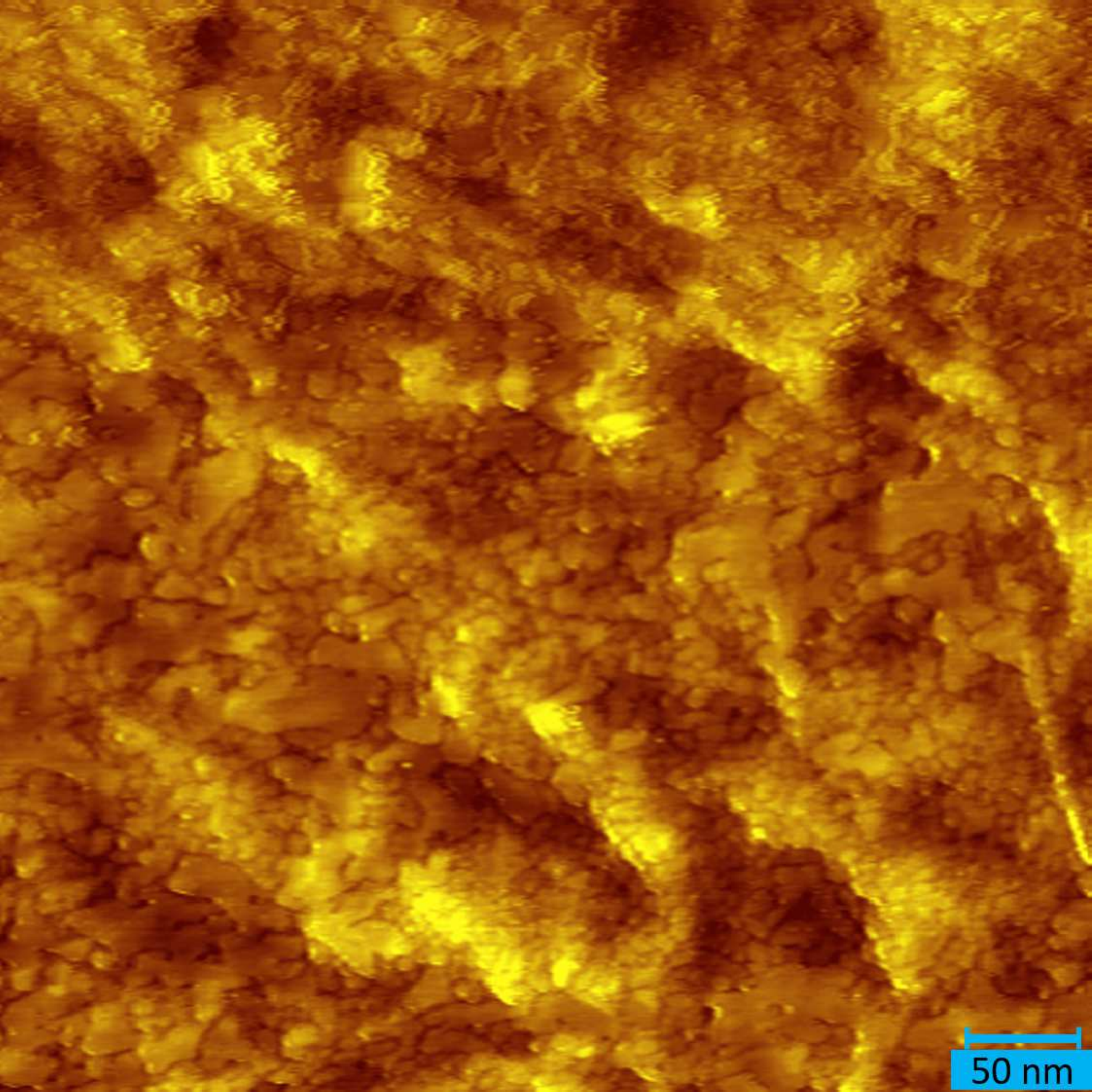}
   \end{center}
   \caption{\label{STM}(a) STM image of the Au layer. Image parameters: V=1.0 V, I=1.0~nA, average RMS roughness: ($0.8\pm0.2$)~nm. (b) STM image of the MLG transferred on the Au layer. Image parameters: V=0.6 V, I=0.5~nA, average RMS roughness: ($1.7\pm0.5$)~nm. The inset shows a cross-sectional plot of a wrinkle taken along the blue line in the STM image. (c) STM image of 12.4 ML Ti evaporated on MLG. Image parameters: V=0.2 V, I=0.09~nA, average RMS roughness: ($2.0\pm0.5$)~nm. All images $500 \times 500$~$\mathrm{nm}^2$.}
\end{figure*}

We characterized the samples after each experimental step, taking STM images of different areas of the Au layer, with the MLG transferred, and after Ti deposition (see Figure~\ref{STM}). The gold surface is corrugated and inhomogeneous, but does not show any wrinkles (see Figure~\ref{STM}(a)). The analysis of the MLG surface shows the typical wrinkles present on graphene on metal substrates after annealing \cite{Obraztsov07} (see Figure~\ref{STM}(b)). In our analysis of the wrinkles we measured heights between 2 and 4~nm, and widths between 10 and 20~nm, in good agreement with results from literature (2--5~nm in height and 5--20~nm in width \cite{Xu09,Zhang11}). From Figure~\ref{STM}(c), taken from a sample with 100\% coverage of titanium, it is clear that Ti atoms cluster and form a layer of large connected islands.

\subsection{Calorimetry during hydrogen exposure}

\begin{figure*}[t]
   \begin{center}
      \includegraphics[width=0.9\textwidth]{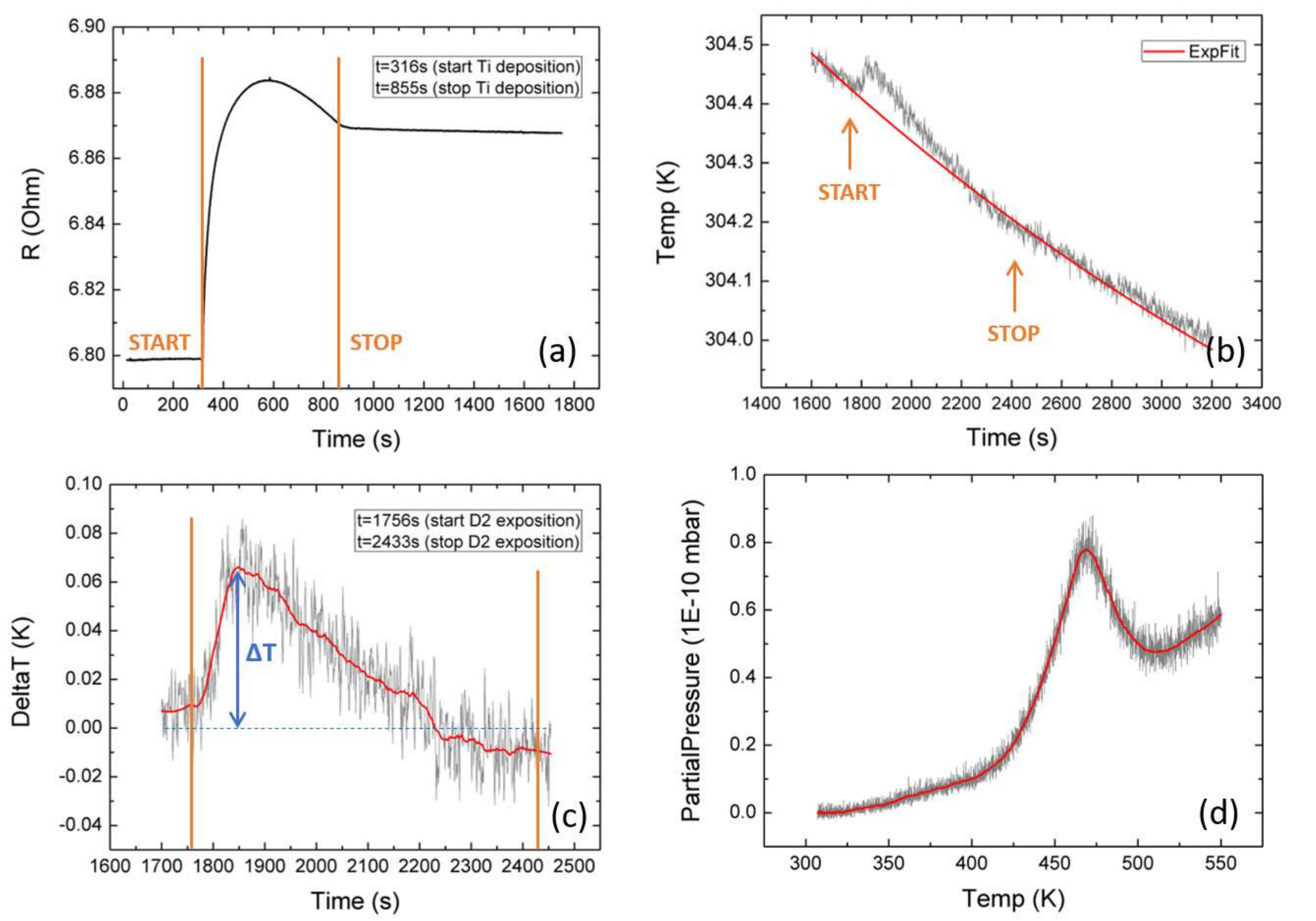}
   \end{center}
   \caption{\label{Ti}(a) Sensor resistance variation during Ti deposition (for 539 s, 12.4 ML of Ti) on MLG. (b) Sensor temperature variation during exposure of the Ti film to D$_2$, showing the increase due to the D$_2$ adsorption process (red line: exponential fit of the thermalization background). (c) Sensor temperature variation during $\mathrm{D_2}$ exposure to the Ti layer (Red line: smoothing). Same data as in (b), with thermalization background subtracted. A $\mathrm{\Delta T=0.065\;K}$ is clearly detected. (d) TDS spectrum of Ti--MLG sample after 5 min of $\mathrm{D_2}$ exposure at $P_{D2}=1.0 \cdot 10^{-7} \;\mathrm{mbar}$. It shows a clear H--desorption peaked at 469 K (Red line: smoothing).}
\end{figure*}

The Au film temperature sensor is so sensitive that the Ti deposition on the MLG (transferred onto the Au sensor) can be followed in real time by recording the sensor resistance. This is shown in Figure~\ref{Ti}(a). As we start the Ti deposition, the resistance increases (by $\sim 0.02-0.08$~$\Omega$) because of the heating of the sample in front of the Ti evaporator, and because of an increased surface diffuse scattering.\cite{Luo94} Subsequently another effect becomes dominant: the Ti layer is acting as an additional resistance in parallel to the Au sensor, and so the effective resultant resistance is decreasing as more Ti is deposited. Once we stop the deposition, the sensor begins to thermalize, and the resistance decreases (following an exponential behavior $T(t)=T_0+A\cdot\exp(-t/\tau)$). Titanium is a very reactive element, and to prevent surface contamination and/or oxidation we do not wait for full thermalization of the sensor after the Ti deposition, but 20 minutes after the Ti deposition we expose the sample to molecular deuterium $(\mathrm{D_2})$ for 5 minutes at a pressure of $1.0 \cdot 10^{-7}\; \mathrm{mbar}$, while recording the sensor resistance. The resulting data are shown in Figure~\ref{Ti}(b). After subtracting the thermalization background (obtained by the exponential fit shown in Figure~\ref{Ti}(b)), we obtain the sensor temperature variation due to D$_2$ adsorption ($\mathrm{\Delta T=0.065\;K}$), as shown in Figure~\ref{Ti}(c). We have verified in control experiments that neither the bare gold sensor nor the graphene film on the sensor display a temperature increase due to exposure to molecular deuterium. This indicates that neither gold nor MLG are able to stably adsorb $\mathrm{D_2}$, in agreement with reports in literature.\cite{Tozzini13} In our experiments we use deuterium ($\mathrm{D_2}$, molecular mass = 4), instead of hydrogen ($\mathrm{H_2}$, molecular mass = 2) in order to obtain a better signal-to-noise ratio in the TDS measurements. 

\subsection{Thermal Desorption Spectroscopy}

Finally, we place the sample in front of a Residual Gas Analyzer (RGA) and heat it at a constant rate of $\sim 0.5$~K/s to a temperature of approximately 600 K. Consistent with previous measurements \cite{Mashoff13,Takahashi16} we observe the desorption of D$_2$ at $T \sim 450$~K. From the measured desorption temperature $T_p=(469\pm3)$ K (extracted from the TDS spectra vs. temperature in Figure~\ref{Ti}(d)) we can estimate the desorption energy barrier $E_d$, which corresponds to the average binding energy per molecule. Using the Redhead equation:\cite{Takahashi16}
\begin{equation}\label{Taka}
E_d/k_b\cdot T_p=A\cdot\tau_m\cdot \exp{\left(-E_d/k_b\cdot T_p\right)}
\end{equation}
where $\tau_m$ is the time from the start of the desorption ramp to the moment at which the desorption peak $T_p$ is reached, $A=10^{13}\;\mathrm{s^{-1}}$, and $k_b=8.625\cdot10^{-5}\;\mathrm{eV\cdot K^{-1}}$ the Boltzmann constant, we obtain $E_d=(1.32\pm0.07)$ eV/molecule. This energy, which corresponds to the average binding energy for the adsorption of deuterium molecules on Ti, is comparable with the ones obtained in previous experimental investigations on similar systems, which were in the range of 1 to 1.5~eV.\cite{Mashoff13,Takahashi16} The amount of desorbed (and therefore stored) $\mathrm{D_2}$ can be estimated from the TDS spectrum (vs. time). At a given pressure, the amount of desorbed gas equals the pumping speed of the vacuum system, here $S = 300$ L/s. After the subtraction of the background, we perform an integration of the TDS spectrum, obtaining the area under the curve, $F=(1.40\pm0.01)\cdot10^{-8}$ mbar$\cdot$s. Then, using $pV=FS=nRT$, with $R=8.314\;\mathrm{J K^{-1} mol^{-1}}$ the gas constant, we obtain $(1.71\pm0.01)\cdot10^{-10}$ mol or $n=(1.03\pm0.01)\cdot10^{14}$ desorbed $\mathrm{D_2}$ molecules. Therefore we can estimate the heat released during the adsorption process: $H_r=n\cdot E_d=(21.8\pm1.3)$~$\mathrm{\mu J}$. As we will show in the following, this value is in good agreement with the one obtained through the calorimetric technique.

\subsection{Discussion}

In order to analyze these data, we can describe the system with a simple thermal model in which the thermometer is heated by the absorption of a thermal power $P(t)$ while at the same time it releases energy by heat losses towards the substrate. Here, the first term corresponds to $\delta H_r/\delta t$, where $H_r$ is the total enthalpy release, while the losses are described by $\lambda$, the heat transfer coefficient. These two contributions are related by the following equation:\cite{Cassettari93}
\begin{equation}\label{eq3}
\delta H_r/\delta t=C\cdot\delta \Delta T(t)/\delta t+\lambda\cdot\Delta T(t).
\end{equation}
Knowing the physical dimensions of the sample, we can calculate the sensor heat capacity $C$. Based on the results of finite element simulations, we include in this calculation the graphene layer, the Au thermometer, and the SiO$_2$ layer underneath (which acts as thermal insulation; see Figures S8 and S9 for a complete description of the sample and the heat capacity calculation). We obtain the total heat capacity $C = (15.0 \pm 0.2) \cdot 10^{-6}$~J/K. Next, we can calculate the heat transfer coefficient as $\lambda = C / \tau$,\cite{Cassettari93} where $C$ is the heat capacity of the sensor and $\tau$ is the characteristic time of cooling of the sample. In order to calculate $\tau$, we heat the sample for a short time with a known thermal power, then we switch off the heating and analyze the cooling process. Fitting the measured curve with an exponential decay due to the thermometer thermalization, we obtain $\tau=(2.9\pm0.6)$ s (see Table S3 for further details).

\begin{figure}[t]
   \begin{center}
      \includegraphics[width=0.49\textwidth]{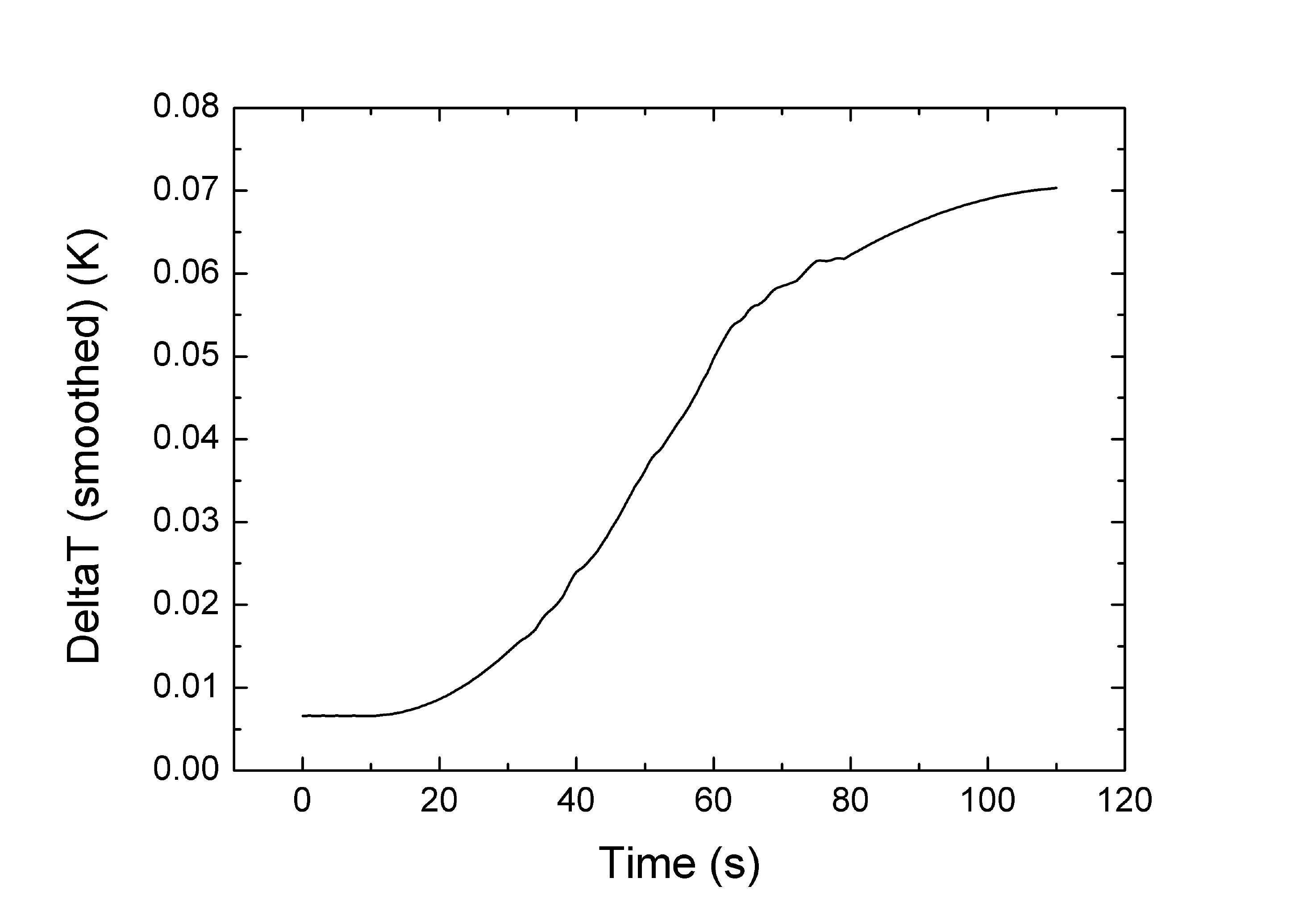}
      \includegraphics[width=0.49\textwidth]{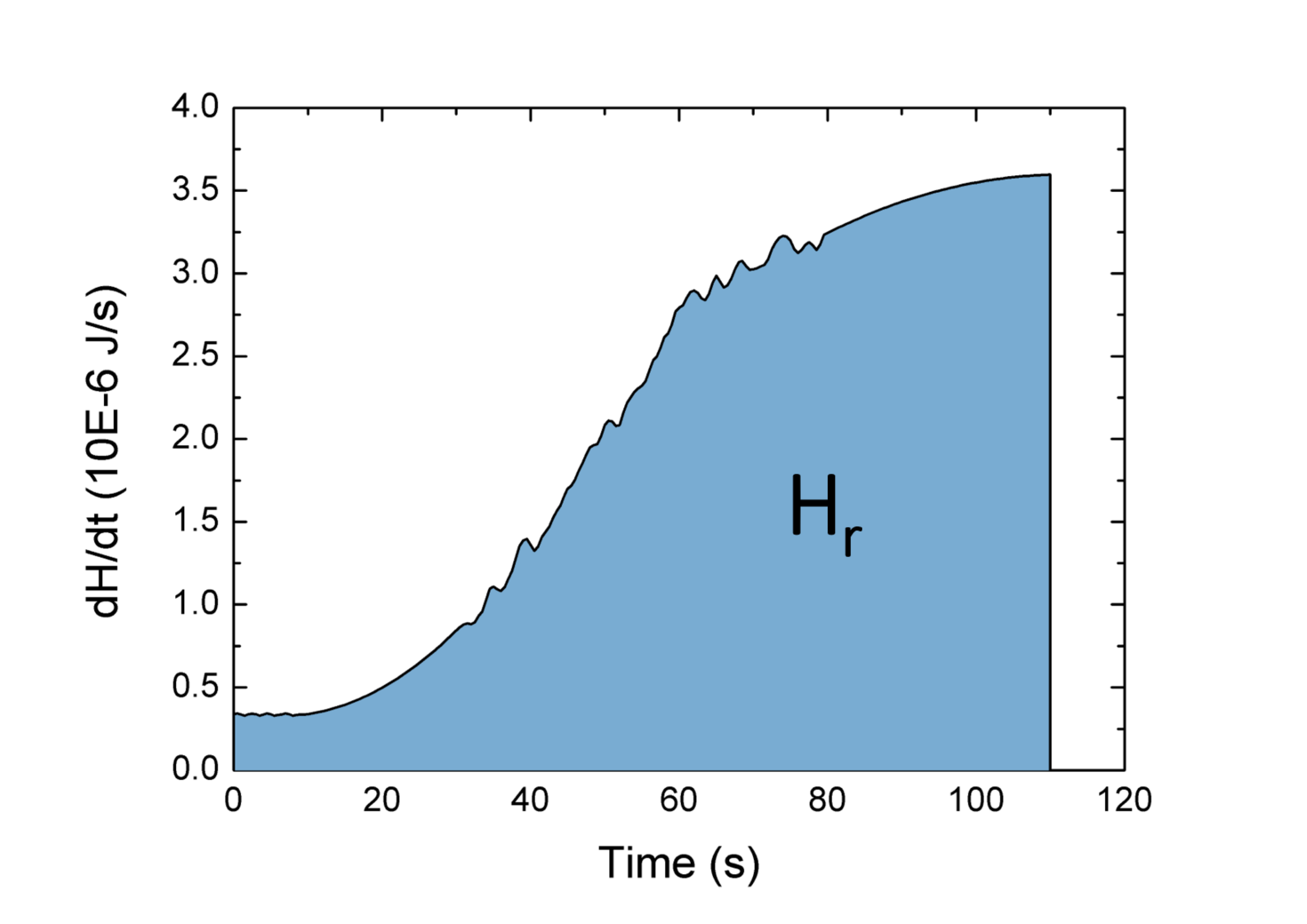}
   \end{center}
   \caption{\label{dDeltatdHr}(a) Rising part of $\Delta T(t)$ (smoothed) at the beginning of $\mathrm{D_2}$ exposure of the Ti--MLG (same data as Figure~\ref{Ti}(c)). (b) $\delta H_r/\delta t$ for the same measurement (the area integration is shown in light blue).}
\end{figure}

This allows us to calculate $\lambda$ as $\lambda = C / \tau = (5.1 \pm 1.1) \cdot 10^{-6}$~W/K. With the values of $C$ and $\lambda$ determined, we can now calculate $\delta H_r / \delta t$ from Eq.~\ref{eq3} using the measured $\Delta T(t)$ curve and performing a point-by-point derivative of the recorded data to obtain $\delta \Delta T(t)/\delta t$. As a final step, we can calculate the enthalpy release $H_r$ from a point-by-point integration of $\delta H_r / \delta t$ and obtain $H_r = (23.4 \pm 4.7)$~$\mu$J. Figure~\ref{dDeltatdHr}(a) shows the rising part of $\Delta T(t)$, while Figure~\ref{dDeltatdHr}(b) shows $\delta H_r / \delta t$, with the area integration which gives the value of $H_r$. The heat release calculated through this calorimetric analysis is in good agreement with the results of the TDS measurement.

\section{Conclusions}

In conclusion, we have developed an original experimental setup allowing the detection of enthalpy release in microscopic devices. With a custom--made $5 \times5$~mm$^2$ sensor we demonstrate a sensitivity of $\sim$~$0.03$~m$\mathrm{\Omega}$ for the resistance variation which corresponds to a sensitivity of $\sim 10$~mK for the temperature variation and $\sim 5$~$\mu$J for the energy release. This calorimetric technique allows to directly measure the enthalpy released during an adsorption process down to $\sim$ 0.2 ng of molecular hydrogen. Besides that, because of its non--destructive, real--time, and scalable characteristics, this innovative method introduces a valuable and reliable new way to measure the amount of adsorbed gas in a solid-state system (as experimentally confirmed by the TDS method). Furthermore, while the calorimetric technique measures \textit{directly} the adsorption process, TDS requires the desorption of the adsorbed gas for the measurement, and is therefore a \textit{post mortem} technique. In particular, we are able to directly detect the small heat release during the exothermic adsorption process of $\sim 10^{-10}$~mol of $\mathrm{D_2}$ on Ti--MLG. Furthermore, the simple procedure for the sample preparation and calibration make this sensor suitable for use in several other applications, where small resistance or temperature variations need to be detected,\cite{Tao17,Hoffmann08} for example ultra-sensitive thermal imaging (with a lower NETD with respect to commercial microbolometers.\cite{Dalsa17,IRnova17,Xenics17}) Moreover, with a proper gold film patterning, this technique could be fruitfully applied to the study of micrometer size samples.

\section{Methods}

\subsection{Fabrication of the gold film thermometer}

A 290~$\mu$m--thick p--type Si(100) wafer polished on the top side was used as a substrate. The top surface of the wafer was covered by a 280~nm--thick SiO$_2$ layer. Metal films (5~nm Ti and 20~nm Au) were deposited on the SiO$_2$ with a SISTEC evaporator. Deposition rate was about 1 \AA/s for both materials. The wafer was successively cut in squares of $5.5 \times 5.5$~mm$^2$. Before graphene transfer, the gold surface was cleaned in a Diener plasma system.

\subsection{CVD growth of graphene}

In this work we used a 25~$\mu$m--thick Cu foil supplied by Alfa Aesar (purity 99.8\%). The foil was electropolished in an electrochemical cell made using a commercially available Coplin staining jar as vessel and an electrolyte solution as described previously.\cite{Miseikis15} Polycrystalline graphene films were synthesized at a pressure of 25 mbar inside a 4 inch cold-wall CVD system (Aixtron HT-BM). The annealing as well as the growth were performed at about 1070 $^{\circ}$C. The temperature was calibrated using the melting point of Cu. The annealing was carried on for 10 minutes in an atmosphere of 1000 sccm of Ar. Growth was performed in an atmosphere of Ar (900 sccm) and H$_2$ (20 sccm), with CH$_4$ (10 sccm) used as a precursor for 20 min. Once the growth process was finished, the system was cooled down in an atmosphere of Ar.

\subsection{PMMA assisted transfer process}

We used a standard PMMA--assisted transfer process for transferring the graphene onto gold--coated SiO$_2$/Si substrates. The graphene/Cu sample was first cleaned in Acetone and Isopropanol (IPA). Once the sample was dry, the top side was spin coated with PMMA for 1 min at 2000 rpm and then kept drying for 30 min. The back layer of graphene on copper was removed by dipping it in a high concentration FeCl$_3$ solution and rinsed several times in DI H$_2$O. The Cu was etched in a 1 mM FeCl$_3$ solution overnight and the PMMA/Gr stack was thoroughly rinsed in DI H$_2$0 and transferred on the target substrate. Finally, the stack was kept in acetone for 2hrs to remove the PMMA, rinsed in Isopropanol, and dried in ambient condition.

\subsection{Characterization}

The sample is characterized by Raman spectroscopy to confirm the quality of graphene (for more details, see the Supporting Information). In order to thermally decouple the sample from the sample holder, we do not use metallic foils for direct current heating of the Si substrate, but a tungsten filament of $\sim 1$~$\mathrm{\Omega}$ resistance integrated in the sample holder for radiative heating. This setup limits the heating rate during TDS to a maximum of $\sim 0.5$~K/s.

\begin{acknowledgments}
We thank Fabio Beltram for his continuous support. Funding from the European Union Horizon 2020 research and innovation programme under Grant Agreement No. 696656 Graphene Core 1 is acknowledged. Financial support from the Consiglio Nazionale delle Ricerche (CNR) in the framework of the agreements on scientific collaborations between CNR and CNRS (France), National Research Foundation of Korea, and RFBR (Russia) is also acknowledged. We further acknowledge funding from the Italian Ministry of Foreign Affairs, Direzione Generale per la Promozione del Sistema Paese, and from the Polish Ministry of Science and Higher Education, Department of International Cooperation (agreement on scientific collaboration between Italy and Poland). We also acknowledge support from Scuola Normale Superiore, Project SNS16\_B\_HEUN 004155.
\end{acknowledgments}

\bibliography{LucaBib}

\end{document}